\documentclass[12pt,onecolumn]{IEEEtran}
\usepackage[colorlinks,bookmarksopen,bookmarksnumbered,citecolor=red,urlcolor=red,]{hyperref}
 \usepackage{indentfirst}
\usepackage{algorithm,algorithmic,amsbsy,amsmath,amssymb,epsfig,bbm,mathrsfs, bbm} 
\usepackage{multirow}
\usepackage{mathrsfs}
 \usepackage{epstopdf}
\usepackage[subfigure]{graphfig}

\usepackage{verbatim}
\usepackage{amsfonts}
\usepackage{amsthm}
\begin{document}

\title{Wireless Charger Networking for Mobile Devices: Fundamentals, Standards, and Applications}

\author{Xiao Lu, Dusit Niyato, Ping Wang, Dong In Kim, and Zhu Han
\thanks{\textbf{D. Niyato} is the corresponding author (Email: dniyato@ntu.edu.sg). } }

\maketitle

\begin{abstract}
 
Wireless charging is a technique of transmitting power through an air gap to an electrical device for the purpose of energy replenishment. Recently, the wireless charging technology has been significantly advanced in terms of efficiency and functionality. This article first presents an overview and fundamentals of wireless charging. We then provide the review of standards, i.e., Qi and Alliance for Wireless Power (A4WP), and highlight on their communication protocols. Next, we propose a novel concept of wireless charger networking which allows chargers to be connected to facilitate information collection and control. We demonstrate the application of the wireless charger network in user-charger assignment, which clearly shows the benefit in terms of reduced costs for users to identify the best chargers to replenish energy for their mobile devices.
 
\end{abstract}

\section{Introduction}

Wireless charging technology enables wireless power transfer from a power source (e.g., a charger) to a load (e.g., a mobile device) across an air gap. The technology provides convenience and better user experience. Recently, wireless charging is rapidly evolving from theories towards standards, and adopted in commercial products, especially mobile phones and portable devices. Using wireless charging has many benefits. Firstly, it improves user-friendliness as the hassle from connecting cables is removed. Different brands and different models of devices can also use the same charger. Secondly, it provides better product durability (e.g., waterproof and dustproof) for contact-free devices. Thirdly, it enhances flexibility, especially for the devices that replacing their batteries or connecting cable for charging is costly, hazardous, or infeasible (e.g., body-implanted sensors). Fourthly, wireless charging can provide on-demand power, avoiding an overcharging problem and minimizing energy costs. In 2014, many leading smartphone manufacturers, e.g., Samsung, Apple and Huawei, release their products equipped with built-in wireless charging capability. IMS Research (www.imsresearch.com) envisioned that wireless charging will have a $4.5$ billion market by 2016. Pike Research (www.pikeresearch.com) estimated that wireless powered products will be tripled by 2020 to a $15$ billion market.

In this article, we first describe a brief history of wireless power transfer technologies. Then, we present an overview and fundamentals of wireless charging technologies. This is then followed by an introduction of two leading international wireless charging standards, i.e., Qi and Alliance for Wireless Power (A4WP). We describe data communication protocols used in these standards. Furthermore, we find that the existing standards mainly focus on the data communication between a charging device and charger, and overlook the communication among chargers and other entities as a network. Therefore, we propose the concept of wireless charger networking to facilitate data communication and information transfer functions among the chargers. We demonstrate the application of the wireless charger network through the user-charger assignment problem. With the wireless charger networking, the cost of user-charger assignment can be minimized.


\section{Overview of Wireless Charging Technique}
\label{sec:overview}




Nikola Tesla, the founder of alternating current electricity, was the first to conduct experiment of wireless charging. He achieved a major breakthrough in 1899 by transmitting $10^8$ volts of high-frequency electric power over a distance of 25 miles to light 200 bulbs and run an electric motor. In 1901, Tesla constructed the Wardenclyffe Tower to transfer electrical energy globally without cords through the Ionosphere. However, due to technology limitation (e.g., low system efficiency), the idea has not been widely further developed and commercialized. Later, during 1920s and 1930s, magnetrons were invented to convert electricity into microwaves, which enables wireless power transfer over long distance. However, there was no method to convert microwaves back to electricity until 1964, when W. C. Brown realized this through a rectenna. Brown demonstrated the practicality of microwave power transfer by powering a model helicopter, which inspired a series of research in microwave-powered airplanes during 1980s and 1990s in Japan and Canada~\cite{J.1988Schlesak}. More recently, different consortiums, e.g., Wireless Power Consortium~\cite{WPC}, Power Matters Alliance~\cite{PMA}, and Alliance for Wireless Power~\cite{WiPower}, have been established to develop international standards for wireless charging. Nowadays, the standards are adopted in many products in the market.

\subsection{Wireless Charging Techniques}

\begin{figure} 
\centering
\subfigure [Inductive Coupling] {
 \label{IC}
 \centering
 \includegraphics[width=0.36 \textwidth]{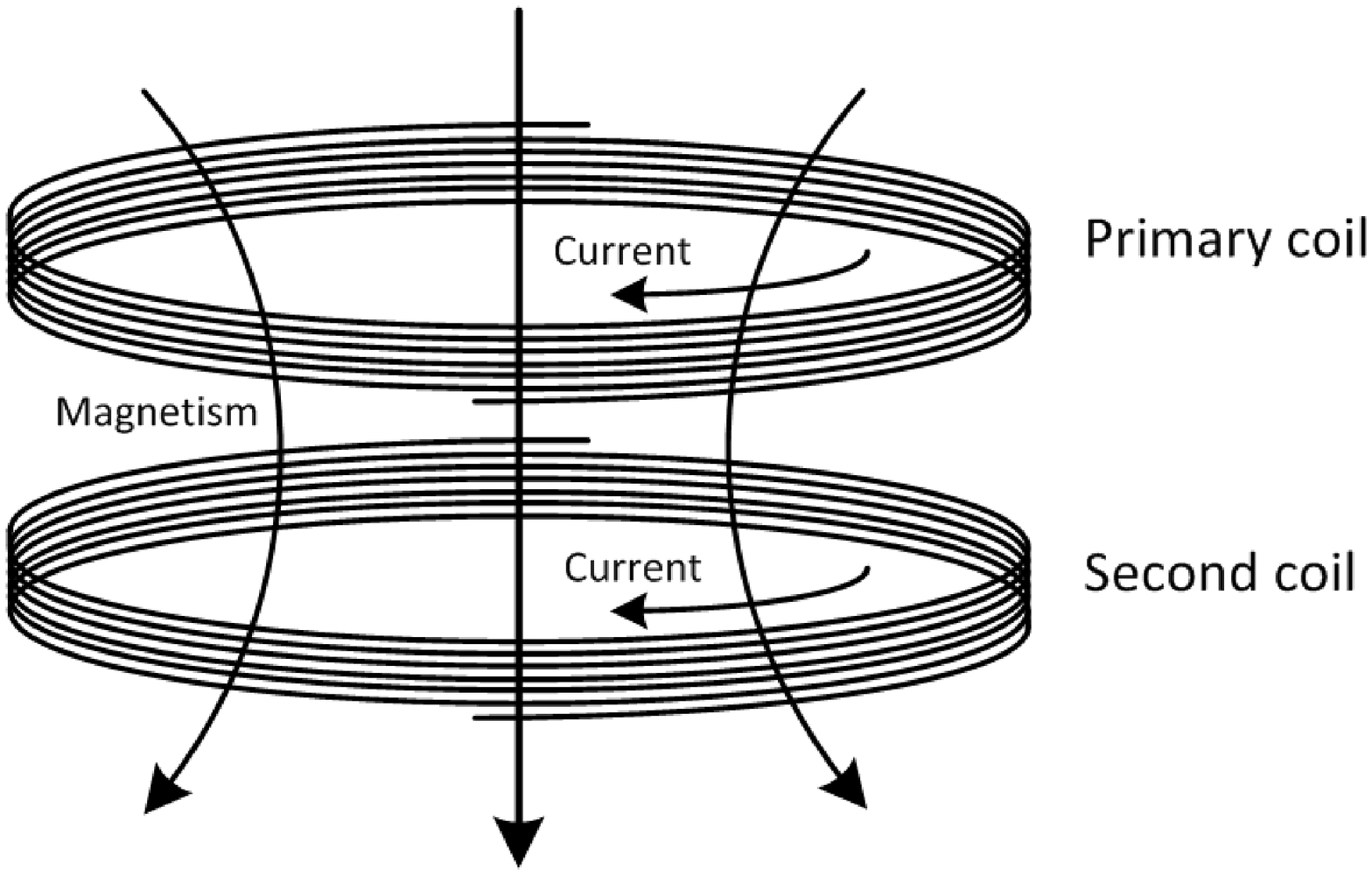}}  
 \centering
 \subfigure [Magnetic Resonance Coupling] {
  \label{MRC}
  \centering
  \includegraphics[width=0.5  \textwidth]{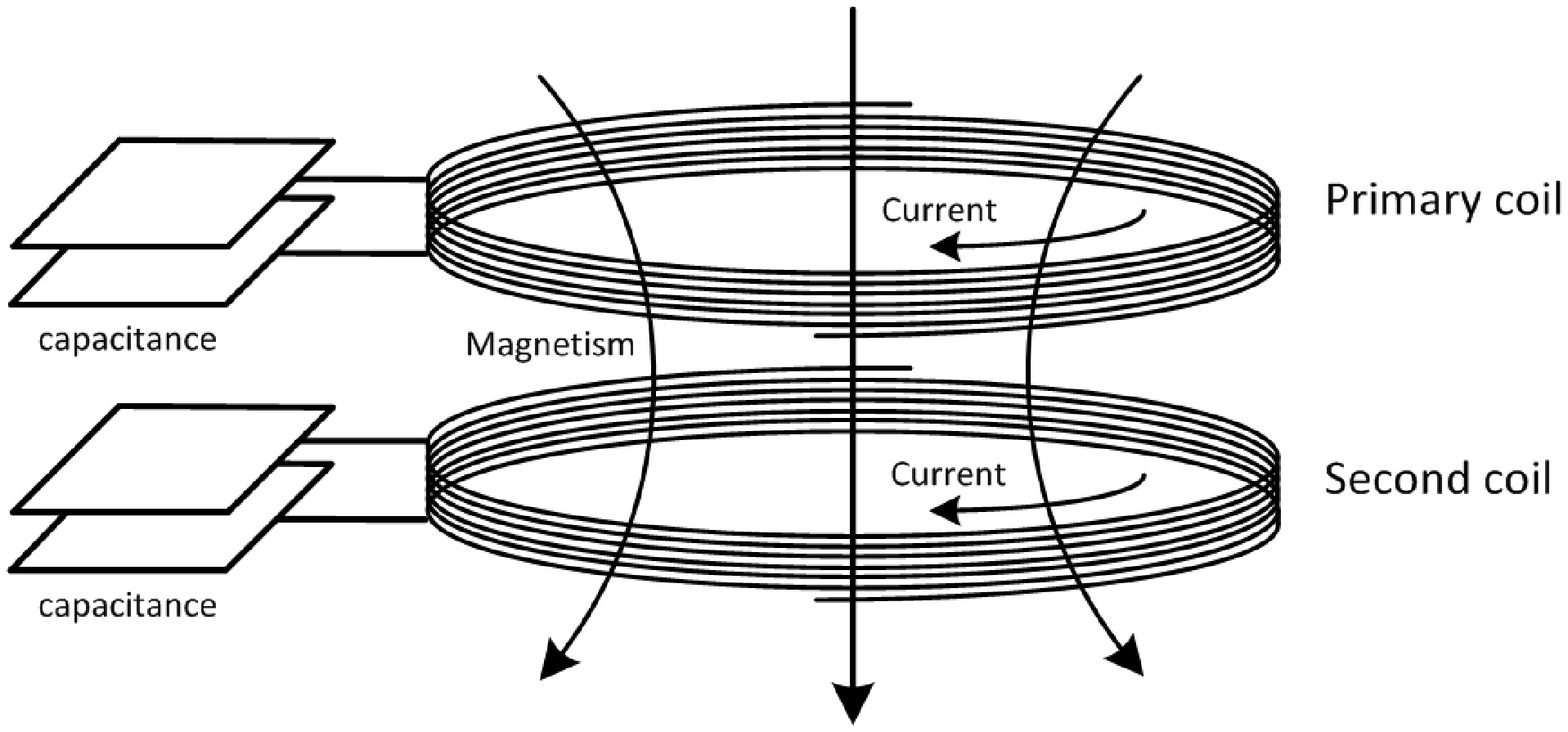}}   
  \centering
 \subfigure [Far-field Wireless Charging] {
 \label{microwave}
  \centering
\includegraphics[width=0.7  \textwidth]{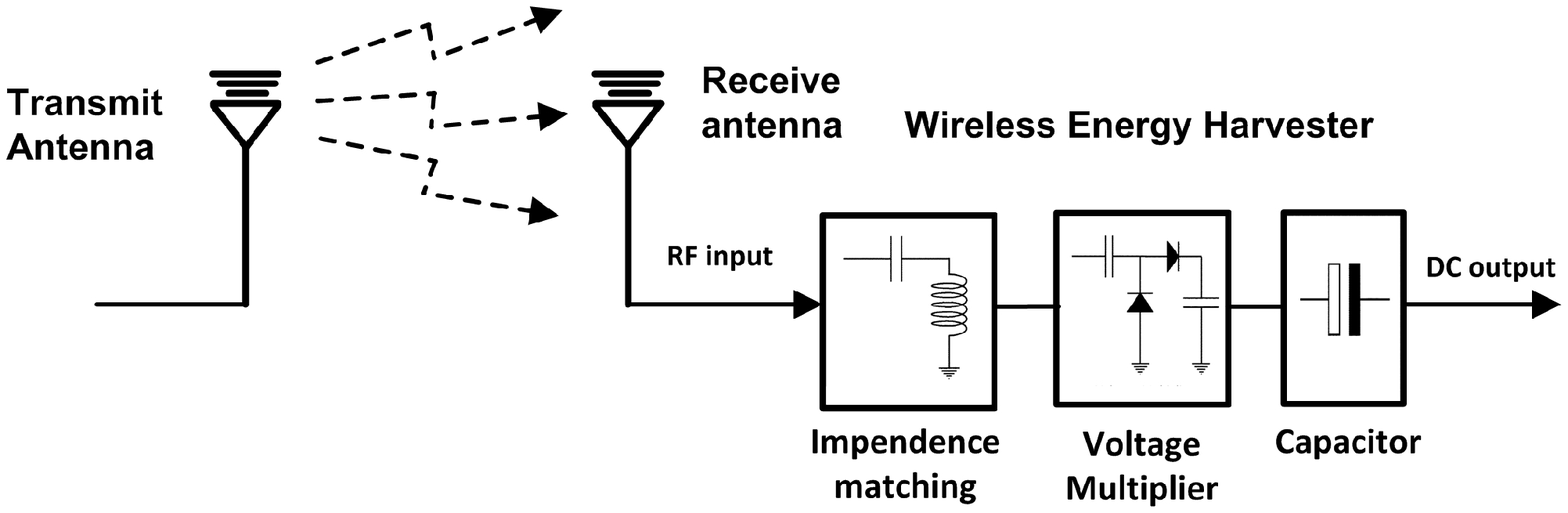}}\\
\centering
\caption{Models of Wireless Charging System.} 
\label{receiver_designs}
\end{figure}

Three major techniques for wireless charging are magnetic inductive coupling, magnetic resonance coupling, and microwave radiation. The magnetic inductive and magnetic resonance coupling work on near field, where the generated electromagnetic field dominates the region close to the transmitter or scattering object. The near-field power is attenuated according to the cube of the reciprocal of the distance. Alternatively, the microwave radiation works on far field at a greater distance. The far-field power decreases according to the reciprocal of the distance. Moreover, for the far-field technique , the absorption of radiation does not affect the transmitter. By contrast, for the near-field techniques, the absorption of radiation influences the load on the transmitter. 

\subsubsection{Magnetic Inductive Coupling}
 
Magnetic inductive coupling~\cite{WPC} is based on magnetic field induction that delivers electrical energy between two coils. Figure~\ref{IC} shows the reference model. Magnetic inductive coupling happens when a primary coil of an energy transmitter generates predominant varying magnetic field across the secondary coil of the energy receiver within the field, generally less than the wavelength. The near-field power then induces voltage/current across the secondary coil of the energy receiver within the field. This voltage can be used by a wireless device. The energy efficiency depends on the tightness of coupling between two coils and their quality factor. The tightness of coupling is determined by the alignment and distance, the ratio of diameters, and the shape of two coils. The quality factor mainly depends on the materials, given the shape and size of the coils as well as the operating frequency. The advantages of magnetic inductive coupling include ease of implementation, convenient operation, high efficiency in close distance (typically less than a coil diameter) and safety. Therefore, it is applicable and popular for mobile devices.
Very recently, MIT scientists have announced the invention of a novel wireless charging technology, called MagMIMO \cite{MagMIMO}, which manages to charge a wireless device from up to 30 centimeters away. It is claimed that MagMIMO can detect and cast a cone of energy towards a phone, even when the phone is put inside the pocket.   

\subsubsection{Magnetic Resonant Coupling}

Magnetic resonance coupling~\cite{Kurs2007A}, as shown in Fig.~\ref{MRC}, is based on evanescent-wave coupling which generates and transfers electrical energy between two resonant coils through varying or oscillating magnetic fields. As the resonant coils, operating at the same resonant frequency, are strongly coupled, high energy transfer efficiency can be achieved with small leakage to non-resonant externalities. This property also provides the advantage of immunity to neighboring environment and line-of-sight transfer requirement. Compared to magnetic inductive coupling, another advantage of magnetic resonance charging is longer effective charging distance. Additionally, magnetic resonant coupling can be applied between one transmitting resonator and many receiving resonators, which enables concurrent charging of multiple devices.

In 2007, MIT scientists proposed a high-efficient mid-range wireless power transfer technology, i.e., Witricity, based on strongly coupled magnetic resonance. It was reported that wireless power transmission can light a $60W$ bulb in more than two meters with transmission efficiency around $40\%$~\cite{Kurs2007A}. The efficiency increased up to $90\%$ when the transmission distance is one meter. However, it is difficult to reduce the size of a Witricity receiver because it requires a distributed capacitive of coil to operate. This poses big challenge in implementing Witricity technology in portable devices. Resonant magnetic coupling can charge multiple devices concurrently by tuning coupled resonators of multiple receiving coils~\cite{A.2010Kurs}. This has been shown to achieve improved overall efficiency. However, mutual coupling of receiving coils can result in interference, and thus proper tuning is required.

\subsubsection{Microwave Radiation}

Microwave radiation \cite{X.NetworkLu} utilizes microwave as a medium to carry radiant energy. Microwaves propagate over space at the speed of light, normally in line-of-sight. Figure~\ref{microwave} shows the architecture of a microwave power transmission system. The power transmission starts with the AC-to-DC conversion, followed by a DC-to-RF conversion through magnetron at the transmitter side. After propagated through the air, the microwaves captured by the receiver rectenna are rectified into electricity again. The typical frequency of microwaves ranges from $300MHz$ to $300GHz$. The energy transfer can use other electromagnetic waves such as infrared and X-rays. However, due to safety issue, they are not widely used.

The microwave energy can be radiated isotropically or towards some direction through beamforming. The former is more suitable for broadcast applications. For point-to-point transmission, beamforming transmit electromagnetic waves, referred to as power beamforming~\cite{ZhangRuiMIMO}, can improve the power transmission efficiency. A beam can be generated through an antenna array (or aperture antenna). The sharpness of power beamforming improves with the number of transmit antennas. The use of massive antenna arrays can increase the sharpness. The recent development has also brought commercial products into the market. For example, the Powercaster transmitter and Powerharvester receiver~\cite{Powercast} allow 1W or 3W isotropic wireless power transfer.

Besides longer transmission distance, microwave radiation offers the advantage of compatibility with existing communication system. Microwaves have been advocated to deliver energy and transfer information at the same time~\cite{Varshney2008}. The amplitude and phase of microwave are used to modulate information, while the radiation and vibration of microwaves are used to carry energy. This concept is referred to as simultaneous wireless information and power transfer (SWIPT)~\cite{ZhangRuiMIMO}. However, due to health concern of RF radiations, the power beacons are constrained by the Federal Communications Commission (FCC) regulation, which allows up to 4 watts for effective isotropic radiated power, i.e., 1 watt device output power plus 6dBi of antenna gain. Therefore, dense deployment of power beacons is required to power hand-held cellular mobiles with lower power and shorter distance. 
The microwave energy harvesting efficiency is significantly dependent on the power density at receive antenna. A detailed survey of energy harvesting efficiency performance of state-of-the-art hardware designs can be found in Table III of \cite{X.1406.6470Lu}.

Table~\ref{WET} shows a summary of the wireless charging techniques. The advantage, disadvantage, effective charging distance and applications are highlighted.

\begin{table}
\footnotesize
\centering
\caption{\footnotesize Comparison of different wireless charging techniques.} \label{WET}
\begin{tabular}{|p{2.5cm}|p{3.5cm}|p{4cm}|p{2.5cm}| p{3cm}|} 
\hline
\footnotesize Wireless charging technique& Advantage & Disadvantage & Effective charging distance   & Applications \\
\hline
Inductive coupling & Safe for human, simple implementation  & Short charging distance, heating effect, Not suitable for mobile applications, needs tight alignment between chargers and charging devices & From a few millimeters to a few centimeters &  Mobile electronics (e.g, smart phones and tablets), toothbrush, RFID tags, contactless smart cards \\
\hline
Magnetic resonance coupling & Loose alignment between chargers and charging devices, charging multiple devices simultaneously on different power, High charging efficiency, Non-line-of-sight charging  & Not suitable for mobile applications, Limited charging distance, Complex implementation &  From a few centimeters to a few meters & Mobile electronics, home appliances (e.g., TV and desktop), electric vehicle charging \\ 
\hline
Microwave radiation & Long effective charging distance, Suitable for mobile applications & Not safe when the RF density exposure is high, Low charging efficiency, Line-of-sight charging  & Typically within several tens of meters, up to several kilometers &  RFID cards, wireless sensors, implanted body devices, LEDs \\
\hline
\end{tabular}
\end{table}

\subsection{Standards}
 
Different wireless charging standards have been proposed. Among them, Qi and A4WP are two leading standards supported by major smartphone manufacturers. This subsection presents an overview of these two standards. 

\begin{figure} 
\centering
\subfigure [Qi-compliant wireless power transfer model] {
 \label{WPT_Qi}
 \centering
 \includegraphics[width=0.8 \textwidth]{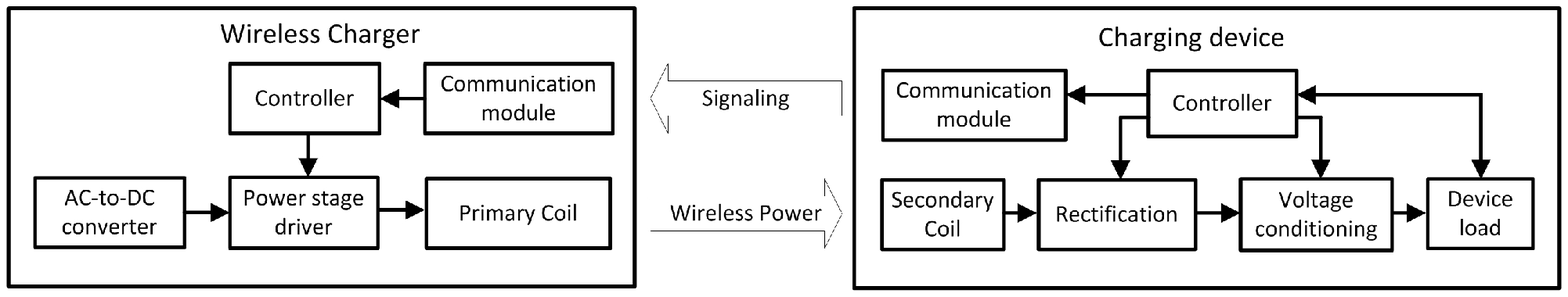}}  
 \centering
 \subfigure [A4WP-compliant wireless power transfer model] {
  \label{WPT_A4WP}
  \centering
  \includegraphics[width=0.8 \textwidth]{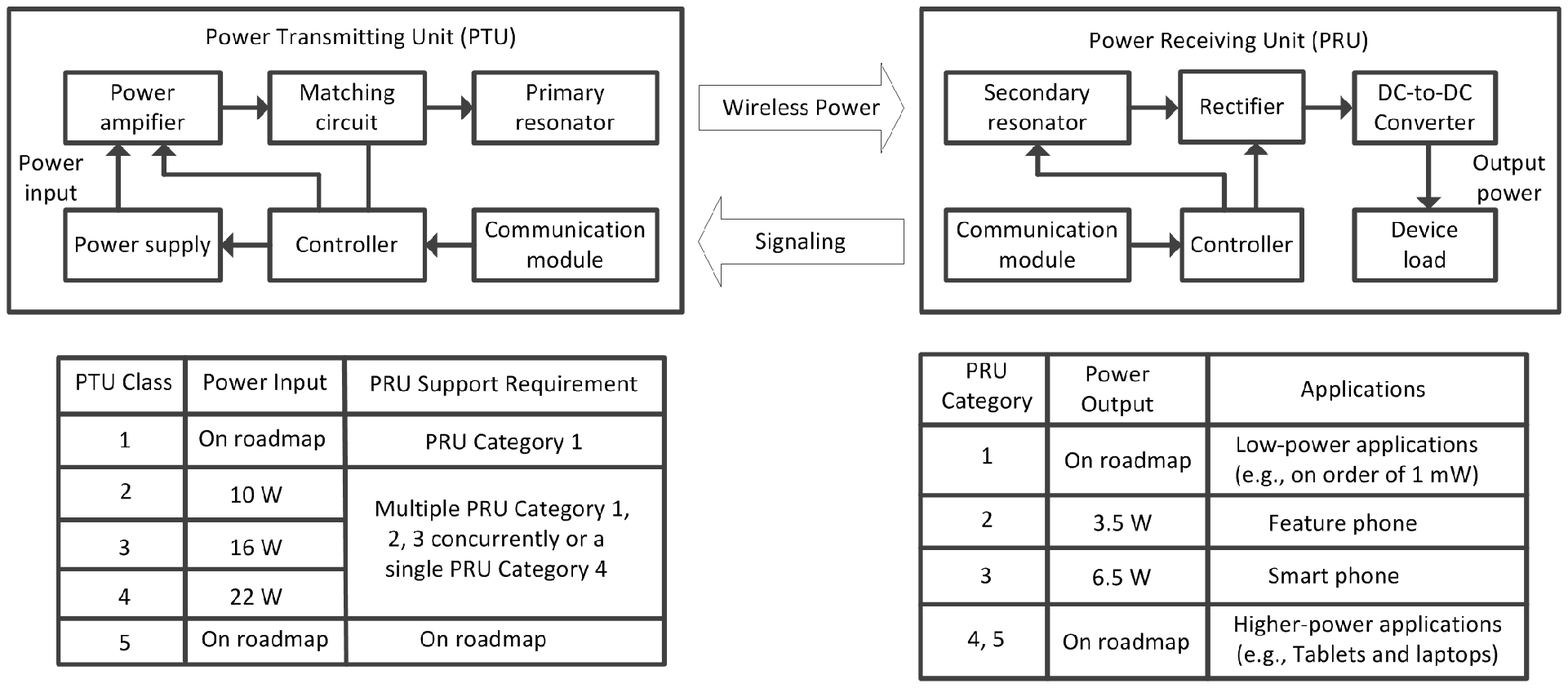}}   
  \centering
\caption{Reference Models of Near-field Wireless Power Transfer Protocol.} 
\label{WPT}
\end{figure}

\subsubsection{Qi}

Qi (pronounced ``chee'') is a wireless charging standard developed by Wireless Power Consortium (WPC)~\cite{WPC}. A typical Qi-compliant system model is illustrated in Fig.~\ref{WPT_Qi}. Qi standard specifies interoperable wireless power transfer and data communication between a wireless charger and a charging device. Qi allows the charging device to be in control of the charging procedure. The Qi-compliant charger is capable of adjusting the transmit power density as requested by the charging device through signaling. 

Qi uses the magnetic inductive coupling technique, typically within the range of 40 millimetres. Two categories of power requirement are specified for Qi wireless charger: 
\begin{itemize}
	\item Low-power category which can transfer power within 5W on 110 to 205 kHz frequency range, and
	\item Medium-power category which can deliver power up to 120W on 80-300 kHz frequency range.
\end{itemize}
Generally, a Qi wireless charger has a flat surface, referred to as a charging pad, which a mobile device can be laid on top. As aforementioned, the tightness of coupling is a crucial factor in inductive charging efficiency. To achieve tight coupling, a mobile device must be strictly placed in proper alignment with the charger. Qi specifies three different approaches for making alignment.
\begin{itemize}
	\item Guided positioning, i.e., a one-to-one fixed-positioning charging, provides guideline for a charging device to be placed, for attaining an accurate alignment. The Qi specification achieves this by using a magnetic attractor. This approach is simple; however, it may require implementation of a piece of material attracted by a magnet in the charging device.   

	\item Free-positioning with movable primary coil, is also a one-to-one charging that can locate the charging device. This approach requires a mechanically movable primary coil that tunes its position to make coupling with the charging device.

	\item Free-positioning with coil array, allows multiple devices to be charged simultaneously irrespective of their positions. This approach can be applied based on the three-layer coil array structure~\cite{R.2005Hui}. Though offering the advantage of user-friendliness, this approach incurs more implementation cost. 
\end{itemize}

The Qi-compliant wireless charging model supports in-band communication. The data transmission is on the same frequency band as that used for the wireless charging. The Qi communication and control protocol is defined to enable a Qi wireless charger to adjust its power output for meeting the demand of the charging device and to disable power transfer when charging is finished. The protocol works as follows.
\begin{itemize}

\item Start: A charger senses the presence of a potential charging devices.  

\item Ping: The charging device informs the charger the received signal strength, and the charger detects the response.  

\item Identification and Configuration: The charging device indicates its identifier and required power while the charger configures energy transfer.
 
\item Power Transfer: The charging device feeds back the control data, based on which the charger performs energy transfer. 

\end{itemize}

 
\subsubsection{Alliance for Wireless Power (A4WP)} 

A4WP aims to provide spatial freedom for wireless power~\cite{R.2013Tseng}. This standard proposes to generate a larger electromagnetic field with magnetic resonance coupling. To achieve spatial freedom, A4WP standard does not require precise alignment and even allows separation between a charger and charging devices. The maximum charging distance is up to several meters. Moreover, multiple devices can be charged concurrently with different power requirement. Another advantage of A4WP over Qi is that foreign objects can be placed on an operating A4WP charger without causing any adverse effect. Therefore, the A4WP charger can be embedded in any object, improving the flexibility of charger deployment.

Figure~\ref{WPT_A4WP} shows the reference model for A4WP-compliant wireless charging. It consists of two components, i.e., power transmitter unit (PTU) and  power receiving unit (PRU). The wireless power is transferred from PTU to PRU, which is controlled by a charging management protocol. Feedback signaling is performed from PRU to PTU to help control the charging. The wireless power is generated at 6.78 MHz ISM frequency. Unlike Qi, out-of-band communication for control signaling is adopted and operates at 2.4 GHz, i.e., typical ISM frequency utilized for near-field communication.

\begin{itemize}

\item A PTU, or A4WP charger has three main functional units, i.e., resonator and matching circuit components, power conversion components, and signaling and control components. The PTU can be in one of following function states: {\em Configuration}, at which PTU does a self-check; {\em PTU Power Save}, at which PTU periodically detects changes of impedance of the primary resonator; {\em PTU Low Power}, PTU establishes a data connection with PRU(s); {\em PTU Power Transfer}, which is for regulating power transfer; {\em Local Fault State}, which happens when the PTU experiences any local fault conditions such as over-temperature; and {\em  PTU Latching Fault}, which happens when rogue objects are detected, when a system error or other failures are reported.
 
\item The A4WP PRU comprises the components for energy reception and conversion, control and communication. The PRU has the following functional states: {\em Null State}, where the PRU is under voltage; {\em PRU Boot}, when the PRU establishes a communication link with the PTU, {\em PRU On}, the communication is performed; {\em PRU System Error State}, when there is over-voltage, over-current, or over-temperature alert; {\em PRU System Error}, when there is an error that has to shut down the power. 
\end{itemize} 
Figure~\ref{WPT_A4WP} also shows the classes and categories for the PTU and PRU (e.g., for power input and output, respectively). No power more than that specified shall be drown for both PTU and PRU. 
 
Similar to Qi standard, A4WP also specifies a communication protocol to support wireless charging functionality. A4WP-compliant systems adopt a Bluetooth Low Energy (BLE) link for the control of power levels, identification of valid loads, and protection of non-compliant devices. The A4WP communication protocol has three steps.
\begin{itemize}
\item Device detection: The PRU that needs to be charged sends out advertisements. The PTU replies with a connection request after receiving any advertisement. Upon receiving any connection request, the PRU stops sending advertisements. Then, a connection is established between the PTU and PRU. 

\item Information exchange: The PTU and PRU exchange their {\em Static Parameters} and {\em Dynamic Parameters} as follows. First, the PTU receives and reads the information of the PRU {\em Static Parameters} which contains its status. Then, the PTU specifies its capabilities in the PTU {\em Static Parameters} and sends them to the PRU. The PTU receives and reads the PRU {\em Dynamic Parameters} that include PRU current, voltage, temperature, and functional status. The PTU then indicates in the {\em PRU Control} to manage charging process.

\item Charging control: It is initiated when {\em PRU Control} is specified and the PTU has enough power to meet the PRU's demand. The PRU {\em Dynamic Parameter} is updated periodically to inform the PTU with the latest information so that the PTU can adjust {\em PRU Control} accordingly. If a system error or complete charging event is detected, the PRU sends PRU alert notifications to the PTU. The PRU {\em Dynamic Parameter} includes the reason for the alert.
\end{itemize}

Apart from the data communication between a charger and charging device, all of the standards do not support information exchange among multiple chargers. Nevertheless, such information exchange can improve the usability and efficiency of the chargers. In the next section, we will introduce the idea of wireless charger network to serve this purpose.

%


\section{Wireless Charger Networking}

\label{sec:charger_networking}

We introduce the concept of wireless charger networking where the chargers not only can communicate with the charging devices, but also can exchange and transfer information with a server. We first present the architecture and features of the wireless charger network. Then, we focus on the user-charger assignment problem, and demonstrate that the wireless charger network can help reduce energy replenishment cost for the users.

\subsection{Wireless Charger Network}

We present a novel idea of a wireless charger network. The network allows multiple chargers to communicate and exchange information with a server (Fig.~\ref{fig:network}). Such information include availability, location, charging status, and cost of different chargers. Collecting this information, the server can optimize the use of chargers for certain purposes. One of them is the user-charger assignment, which allows the matching between users, who needed to replenish energy for their mobile devices, and available chargers. With the location, availability, and status of each charger, the server can inform the users the best chargers, e.g., the nearest available ones. This information of the chargers has to be updated continuously so that the assignment can be performed in an online fashion with up-to-date information. Therefore, the wireless charger network can serve this purpose.

\subsection{Architecture}

\begin{figure}[h]
\begin{center}
$\begin{array}{c} \epsfxsize=6 in \epsffile{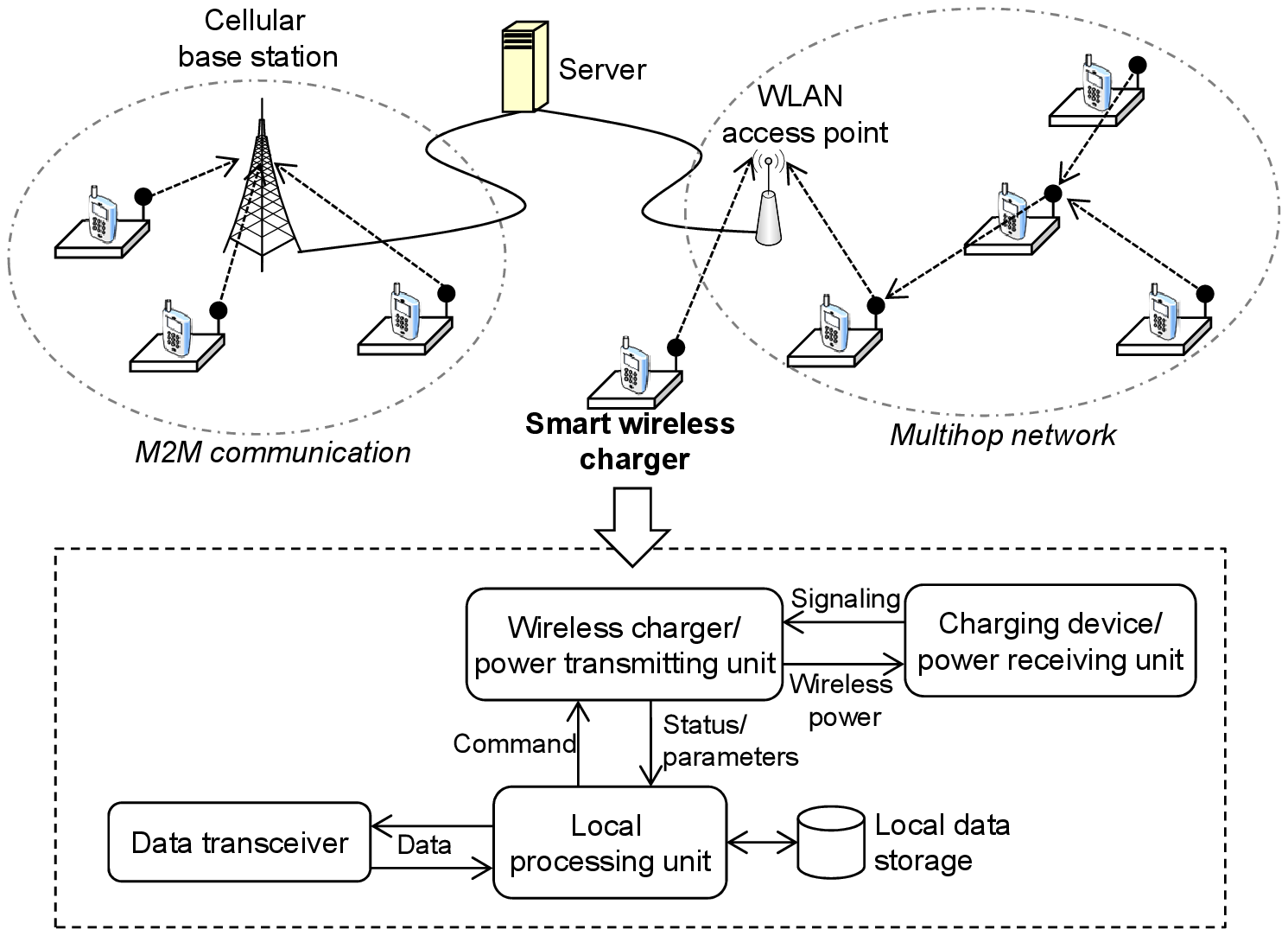} \\ [-0.2cm]
\end{array}$
\caption{An architecture for the wireless charger network.} 
\label{fig:network}
\end{center}
\end{figure}

The major components of the wireless charger network are as follows.

\begin{itemize}
	\item {\em Smart wireless charger:} In addition to typical charging functionality, the smart wireless charger will be equipped with a data transceiver (Fig.~\ref{fig:network}). It has a local processing unit to process and data storage to store data of the wireless chargers (e.g., network setting). It can process and store data from charging devices (e.g., usage statistics and history) and commands from other components in the network (e.g., a status query).
	\item {\em Wireless access point:} This is a typical wireless base station. It can be WLAN access point or cellular base station to provide communication channels to the smart wireless charger. The communication can be a WLAN connection (e.g., WiFi or Bluetooth) or machine-to-machine (M2M) or machine-type communication (MTC) for a cellular connection (e.g., LTE). Moreover, connections among chargers are also possible through multihop networks (e.g., mesh networks). For example, the chargers outside the transmission range of the wireless access point can have their information relayed by intermediate chargers.
	\item {\em Server:} It can perform authentication, authorization and accounting (AAA) and other centralized functions. It has a network data storage for maintaining various information about the wireless charger network (e.g., status and charging price of individual chargers). It also provides an interface with users. For example, the users can contact the server to request for status information of the chargers in the network. The server can also optimize and direct users to the available chargers. 
\end{itemize}

In the wireless charger network, which allows smart wireless chargers to communicate, the following functions can be implemented.
\begin{itemize}
	\item {\em Authentication:} The chargers can verify an identity of a charging device. For example, any chargers in the same network can serve the devices owned by registered users. The account and password information can be exchanged between a charging device and charger. The charger can locally verify the information or can remotely authenticate the device with the server.
	\item {\em Charging payment:} To charge the device, the charger may require payment from a user. The charger can implement different pricing schemes (e.g., time-of-use) programmed by the server. 
	\item {\em Reporting status:} The server can collect information about the chargers in the networks (e.g., location, available charging capacity, energy level and identity of the charging devices). The users who want to use chargers can contact the server to retrieve some of this information (e.g., to choose the best charger). The owner of the wireless charger network can collect usage statistic, e.g., for accounting purpose.
	\item {\em User-charger assignment:} The server can collect information about users required charging services 
	and match them with appropriate chargers to achieve a certain objective. We will demonstrate one simple designs later.
	\item {\em Add-on services:} The charger can provide add-on or value added services, e.g., information downloading and content distribution from a server. The services can benefit from short-range transmission between a charger and a charging device.
\end{itemize}

Note that there are some mobile applications that allow users to locate charging facility, e.g., ChargeBox (http://www.chargebox.com/chargebox-app/). However, they do not provide real-time information about chargers. The concept of wireless charger networking can be adopted to enhance the functionality of such applications.

\subsection{User-Charger Assignment}

We demonstrate the advantage of the wireless charger networking through the user-charger assignment problem. In the problem, there are wireless chargers deployed at different locations. With the wireless charger networking, the status and other information of the chargers can be available to the new users (i.e., the users required to charge their devices). The new users use the information to determine the best charger with the minimum cost. Here we consider the following costs.
\begin{itemize}
	\item {\em User effort} is the measure of a user to move from its current location to a target charger.
	\item {\em Price} is the money paid for charging at the target charger. 
	\item {\em Delay} is the time taken for a user to wait until the device is fully charged.
\end{itemize}
We consider two schemes of user-charger assignment, i.e., individual selection and optimal assignment. 
\begin{itemize}
	\item In the individual selection scheme, each new user chooses the charger which yields the minimum individual cost irrespective of the decision of other users. Let $C_j$ be the estimated overall cost of charger $j$. The user $i$ chooses the charger $j^* = \arg \min_j C_j$, where $C_j = w_1 (T_j + t_i)/n_j + w_2 p_j t_i + w_3 D_{i,j}$. $T_j$ is the total amount of energy to be charged for other users at charger $j$, $n_j$ is the charger capacity, and $p_j$ is the price per unit of charging energy of charger $j$. $t_i$ is the amount of energy to be charged for user $i$, and $D_{i,j}$ is the distance between the current location of user $i$ and charger $j$. $w_1$, $w_2$, and $w_3$ are the weights for the delay, price, and user effort costs, respectively.
	\item In the optimal assignment scheme, new users request the server for charging services. The server then assigns the new users periodically to chargers to minimize total costs. Let $x_{i,j}$ be an indicator variable whose value is one if user $i$ is assigned to charger $j$, and zero otherwise. $x_{i,j}$ is called the assignment for all $i$ and $j$. Let $ C_{i,j}$ be an estimated overall cost if user $i$ is assigned to charger $j$. The optimal assignment is the solution that $x^*_{i,j}	=	\min_{x_{i,j} }	\sum_{i} x_{i,j} C_{i,j}$, where $C_{i,j}		=	w_1 (T_j+t_i) /n_j + w_2 p_j t_i + w_3 D_{i,j}$ has a similar meaning to that in the individual selection scheme. In this case, one user will be assigned only to one of the chargers.
\end{itemize}
These schemes can be implemented in the wireless charger network where complete information about chargers and users is available. Additionally, for comparison purpose, we also consider a simple scheme in which the user always chooses the nearest charger (i.e., the nearest scheme) due to lack of chargers' information (e.g., location, status, etc).

To evaluate the performance improvement from the wireless charger network, we simulate a simple scenario of user-charger assignment. A service area (e.g., a campus) is divided into 16 areas in a grid structure. The first column is composed of areas 1, 2, 3, and 4. The second column is composed of areas 5, 6, 7, and 8, and so on. Each area has a charger with the capacity to charge three devices simultaneously. The distance between neighboring chargers is 125 meters. A user's device requires 20 minutes to be fully charged. We assume that all weights are ones. The charging prices for the chargers at areas $j$ is $p_j = 0.25 + j \times 0.08$ cents per minute. That is, the chargers at areas 1 and 16 have the lowest and highest prices, respectively. There are new users in each area required charging with the rate of 6 users per hour. Note that we assume location information of the charger and new user are available. For example, the location of the users is known through their connected wireless access points.

\begin{figure}[h]
\begin{center}
$\begin{array}{c} \epsfxsize=3.8 in \epsffile{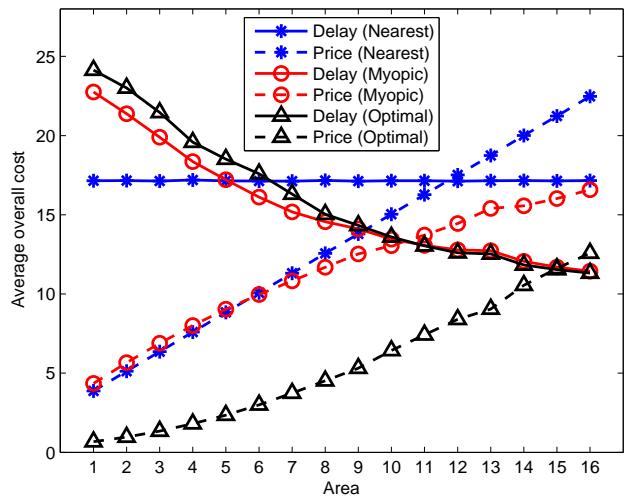} \\ [-0.2cm]
\end{array}$
\caption{Delay and price costs of users at different areas.} 
\label{fig:area}
\end{center}
\end{figure}

Figure~\ref{fig:area} shows the delay and price costs of users charging at different chargers. As the charging price at different location is different, new users choose or are assigned to different chargers accordingly. For the individual selection and optimal assignment schemes, the new users from the areas with higher charging prices (e.g., area 16) will evade to the chargers with lower charging prices (e.g., area 1). Therefore, the number of users using the chargers with low price is higher, causing larger delay. Balancing between charging price and delay, the optimal assignment scheme achieves an average overall cost lower than that of the individual selection scheme (i.e., 22.7 versus 27.9, respectively), and they achieve a much lower cost than that of the nearest scheme, whose average overall cost is 30.3. This results clearly show the benefit of the wireless charger networking in the user-charger assignment problem. 

\begin{figure}[h]
\begin{center}
$\begin{array}{c} \epsfxsize=3.8 in \epsffile{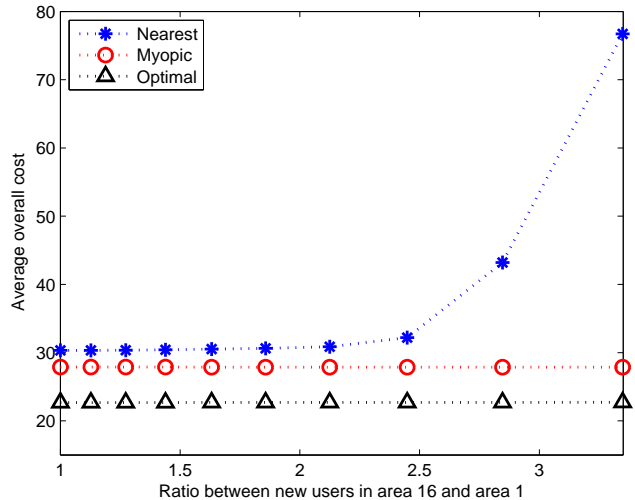} \\ [-0.2cm]
\end{array}$
\caption{Average overall cost when the number of new users at different area is varied.} 
\label{fig:ratio}
\end{center}
\end{figure}

Then, we vary the number of new users (i.e., considered as a load) at different locations, while keeping the total number of new users in all the areas constant. The number of new users increases linearly from areas 1 to 16. Figure~\ref{fig:ratio} shows the average overall cost under different number of new users at different locations (i.e., shown as the ratio between those in areas 16 and 1). The ratio is 1 if all areas have uniform number of new users, and the ratio is 4 if area 16 has more number of new users than that of area 1 for 4 times. We observe that the individual selection and optimal assignment schemes are not affected by the load variation, while the nearest scheme severely experiences the rising overall cost. This is due to the fact that the new users with the individual selection and optimal assignment schemes have access to the information of wireless charger network, providing them an option to select the best charger. The new users observing long waiting time can choose to go to the charger with shorter waiting time. Again, the optimal assignment scheme achieves the lowest overall cost due to the collective user-charger assignment, which relies on the benefits of all users instead of individual users as in the individual selection scheme.


\section{Open Research Issues}
\label{sec:futureissues}

This section highlights some open issues and challenges with both wireless charging technologies and data communication in wireless charging systems.

\subsection{Open Issues in Wireless Charging}

{\em Inductive coupling:} The increase of wireless charging power density gives rise to several technical issues, e.g., thermal, electromagnetic compatibility, and electromagnetic field problems~\cite{Y2013Hui}. This requires high-efficient power conversion techniques to mitigate the power loss at an energy receiver and battery modules with effective ventilation design.

{\em Resonance coupling:} Resonance coupling-based techniques, such as Witritiy and Magnetic MIMO, have a larger charging area and is capable of charging multiple devices simultaneously. However, they also cause increased electromagnetic interference with a lower efficiency compared to inductive charging. Another limitation with resonance coupling is the relatively large size of a transmitter. The wireless charging distance is generally proportional to the diameter of the transmitter. Therefore, wireless charging over long distance typically requires large receiver size.

{\em Magnetic MIMO:} For multi-antenna near-field beamforming, the computation of magnetic-beamforming vector on the transmission side largely depends on the knowledge of the magnetic channels to the receivers. The design of channel estimation and feedback mechanisms is of paramount importance. With the inaccuracy of channel estimation or absence of feedback, the charging performance severely deteriorates. Additionally, there exists a hardware limitation that the impedance matching hardware is optimally operated only within a certain range~\cite{MagMIMO}.

\subsection{Open Issues in Data Communication}

To improve the usability and efficiency of the wireless charger, their data communication capability can be enhanced. 

{\em Duplex communication and multiple access:} The current communication protocols support simplex communication (e.g., from a charging device to charger). However, there are some important procedures which require duplex communication. For example, the charging device can request for a certain charging power, while the charger may request for battery status of the charging device. Moreover, the current protocols support one-to-one communication. However, multiple device charging can be implemented and medium access control (MAC) with multiple access for data transmission among charging devices and charger has to be developed and implemented.

{\em Secure communication:} The current protocols support plain communication between a charger and charging device. They are susceptible for eavesdropping attacks (e.g., to steal charging device's and charger's identity) and man-in-the-middle attacks (e.g., malicious device manipulates or falsifies charging status). The security features have to be developed in the communication protocols, taking unique wireless charging characteristics (e.g., in-band communication in Qi) into account. 

{\em Inter-charger communication:} The protocols support only the communication between a charger and charging device (i.e., intra-charger). Multiple chargers can be connected and their information as well as charging devices' information can be exchanged (i.e., inter-charger). Although Section~\ref{sec:charger_networking} proposes the concept of wireless charger networking, there are some possible improvements. For example, wireless chargers can be integrated with a wireless access point, which is called a hybrid access point, to provide both data communication and energy transfer services.


\section{Conclusion}
\label{sec:conclusion}

Wireless charging technology will become prevalent especially for consumer electronics, mobile, and portable devices. In this article, we have presented an overview and fundamentals of wireless charging techniques. Two major standards, i.e., Qi and A4WP, have been reviewed, with the focus on their data communication protocols. We have discussed about open issues in the protocols. We have then proposed the concept of wireless charger networking to support inter-charger data communication. We have demonstrated its usage for user-charger assignment, which can minimize the cost of users in identifying the best charger to replenish energy of their devices.

\end{document}